\documentclass[10pt]{article}

\usepackage{amsfonts}
\usepackage{amsmath} 
\usepackage{amssymb}

\newcommand{\INC}{}

\usepackage{amsfonts}
\usepackage{amsmath} 
\usepackage{amssymb}
\usepackage{enumerate}

\newcommand{\PROOF}{\begin{proof}}

\newcommand{\QED}{\end{proof}}

\newcommand{\myset}[2]{ \left\{ #1 \left| #2 \right. \right\} }

\newcommand{\prefix}{\sqsubseteq}

\newcommand{\N}{\mathbb{N}}

\newcommand{\Q}{\mathbb{Q}}

\newcommand{\constr}{{\mathrm{constr}}}

\newcommand{\cdim}{\mathrm{cdim}}
\newcommand{\G}{{\mathcal{G}}}
\newcommand{\D}{{\mathcal{D}}}
\newcommand{\Ghat}{\widehat{\G}}

\newcommand{\C}{\mathbf{C}}

\newcommand{\bd}{{\mathbf{d}}}

\usepackage{amsthm}

\newtheorem{theorem}{Theorem}[section]
\newtheorem{corollary}[theorem]{Corollary}
\newtheorem{lemma}[theorem]{Lemma}

\newenvironment{theorem_cite}[1]
{\begin{theorem}  {\rm (#1)}}
{\end{theorem}}

\theoremstyle{definition}

\newtheorem*{definition}{Definition}

\newtheorem*{example*}{Example}
\newtheorem*{examples*}{Examples}

\newtheorem*{remark}{Remark}

\newtheorem*{notation}{Notation}

\newtheorem*{ack}{Acknowledgment}

\theoremstyle{remark}

\numberwithin{equation}{section}
\numberwithin{figure}{section}

\title{
{\bf Gales Suffice for Constructive Dimension}
\footnote{This research was supported in part by National Science Foundation Grant 9988483.}
}

\author{
John M. Hitchcock\\
Department of Computer Science\\
Iowa State University\\
jhitchco@cs.iastate.edu
}

\date{}

\begin{document}

\maketitle

\begin{abstract}
Supergales, generalizations of supermartingales, have been used by
Lutz (2002) to define the constructive dimensions of individual binary
sequences.  Here it is shown that gales, the corresponding
generalizations of martingales, can be equivalently used to define
constructive dimension.
\end{abstract}

\section{Introduction}
Effective martingales have been very useful objects in theoretical
computer science.  Schnorr \cite{Schn71a,Schn71b} used constructive
martingales to give an equivalent definition of Martin-L\"{o}f
randomness \cite{MartinLof66}.  Martingales computable within resource
bounds have been used by Lutz \cite{Lutz:AEHNC} to define various
resource-bounded measures that have been successful in complexity
theory.  In all these cases, it is known that replacing the
constructive or resource-bounded martingales with constructive or
resource-bounded supermartingales results in an equivalent definition.

Lutz \cite{Lutz:DCC} recently introduced supergales and gales as
natural generalizations of supermartingales and martingales,
respectively.  He showed that gales can be used to characterize
classical Hausdorff dimension.  With this as a motivation, Lutz used
gales computable within resource bounds to define resource-bounded
dimensions that work inside of complexity classes.  He also showed
that supergales may be used in place of gales to give equivalent
definitions of these dimensions.

Constructive dimension \cite{Lutz:DISS} refines the theory of
Martin-L\"{o}f randomness by assigning each individual binary sequence
a dimension.  Lutz used constructive supergales to define constructive
dimension.  Supergales were used rather than gales because he was able
to show that optimal constructive supergales exist.  The questions of
whether optimal constructive gales exist and whether gales can be used
to equivalently define constructive dimension were left open.

Therefore, martingales and supermartingales are known to give
equivalent definitions for all the applications mentioned above, and
gales and supergales are known to give equivalent definitions for all
the applications mentioned above {\em except} constructive dimension.
Here it is shown that constructive gales give an equivalent definition
of constructive dimension.  The proof is a simple and direct
construction that uses some ideas from an earlier paper by the author
\cite{Hitchcock:CPED}.  As a corollary we obtain a form of optimal
constructive gales.


\section{Preliminaries}
The set of natural numbers is $\N = \{0,1,2,\ldots\}$.  The set of
binary strings of length $n \in \N$ is $\{0,1\}^n$.  The set of all
finite binary strings is $\{0,1\}^*$.  The empty string is
$\lambda$.  For a language $A \subseteq \{0,1\}^*$, we write $A_{=n}$
for the set of strings in $A$ of length $n$.  For strings $w,v \in
\{0,1\}^*$, we write $w \prefix v$ if $w$ is a prefix of $v$.  $\C$ is
the Cantor space of all infinite binary sequences.  For a sequence $S
\in \C$, $S[0..n-1]$ is the prefix of $S$ of length $n$.

A real number $r$ is computable if there is a computable function $f :
\N \to \Q$ such that $|f(n) - r| \leq 2^{-n}$ for all $n \in \N$.  A
function $g : \{0,1\}^* \to [0,\infty)$ is constructive if there is a
computable function $h : \{0,1\}^* \times \N \to \Q$ such that for all
$w \in \{0,1\}^*$, $h(w,n) \leq h(w,n+1) < g(w)$ for all $n \in \N$ and
$g(w) = \sup_{n\in\N} h(w,n)$.

\section{Constructive Dimension}
Constructive dimension was introduced by Lutz \cite{Lutz:DISS}.  Here
we review the basic concepts.  We begin by defining supergales and
gales.
\begin{definition}
Let $s \in [0,\infty)$.  A function $d : \{0,1\}^* \to [0,\infty)$ is an {\em
$s$-supergale} if
\begin{equation}\label{eq:supergale_cond}
d(w) \geq \frac{d(w0)+d(w1)}{2^s}
\end{equation}
for all $w \in \{0,1\}^*$.  If equality holds in
\eqref{eq:supergale_cond} for all strings $w$, then $d$ is an {\em
$s$-gale}.
\end{definition}

Note that 1-gales are martingales and 1-supergales are
supermartingales.  We are particularly interested in the success sets
of supergales and gales.

\begin{definition}
The {\em success set} of a supergale $d : \{0,1\}^* \to [0,\infty)$ is
$$S^\infty[d] = \myset{S\in\C }{ \limsup_{n\to\infty} d(S[0..n-1]) = \infty
}.$$
\end{definition}

\begin{notation}
For any $X \subseteq \C$, we define the sets 
$$\G_\constr(X) = 
\left\{ s \left| \begin{array}{l}\textrm{there
exists a constructive}\\\textrm{$s$-gale $d$ for which $X
\subseteq S^\infty[d]$}\end{array} \right.\right\}$$
and
$$\Ghat_\constr(X) = 
\left\{ s \left| \begin{array}{l}\textrm{there
exists a constructive}\\\textrm{$s$-supergale $d$ for which $X
\subseteq S^\infty[d]$}\end{array} \right.\right\}$$
of nonnegative real numbers.
\end{notation}

Constructive dimension is defined in terms of succeeding constructive
supergales.
\begin{definition}For a set $X \subseteq \C$, the {\em constructive dimension} of
$X$ is 
$$\cdim(X) = \inf \Ghat_\constr(X).$$
For a sequence $S \in \C$, the {\em constructive dimension} of $S$ is
$$\cdim(S) = \cdim(\{S\}).$$
\end{definition}

We now define two notions of optimality for a class of supergales.
\begin{definition}
Let $d^*$ be a supergale and let $\D$ be a class of supergales.
\begin{enumerate}
\item  We say that $d^*$ is {\em multiplicatively optimal} for $\D$ if
for each $d \in \D$ there is an $\alpha > 0$ such that $d^*(w) \geq
\alpha d(w)$ for all $w \in \{0,1\}^*$.
\item  We say that $d^*$ is {\em successively optimal} for $\D$ if for every
$d \in \D$, $S^\infty[d] \subseteq S^\infty[d^*]$.
\end{enumerate}
\end{definition}

Lutz used Levin's universal constructive semimeasure \cite{ZvoLev70}
to show that there exist multiplicatively optimal supergales.
\begin{theorem_cite}{Lutz \cite{Lutz:DISS}}\label{th:lutz_optimal}For any computable
$s \in [0,\infty)$ there is a constructive $s$-supergale $\bd^{(s)}$
that is multiplicatively optimal for the class of constructive
$s$-supergales.
\end{theorem_cite}
Theorem \ref{th:lutz_optimal} was used to prove the following
cornerstone of constructive dimension theory.
\begin{theorem_cite}{Lutz \cite{Lutz:DISS}}\label{th:lutz_pointwise}For
any $X \subseteq \C$, $$\cdim(X) = \sup_{S \in X} \cdim(S).$$
\end{theorem_cite}

\begin{remark}
In \cite{Lutz:GCDIS}, a conference paper preceding \cite{Lutz:DISS},
Lutz defined constructive dimension using constructive gales.  There
Lutz used a false assertion about martingales to argue that there
exist multiplicatively optimal constructive gales.  These ``optimal
gales'' were then used to prove Theorem \ref{th:lutz_pointwise}.
These flawed arguments were subsequently noticed and corrected in
\cite{Lutz:DISS} by reformulating constructive dimension in terms of
constructive supergales.  The multiplicatively optimal supergales of
Theorem \ref{th:lutz_optimal} exist and Theorem
\ref{th:lutz_pointwise} is true in the reformulation.  However, Lutz
left open the questions of whether there exist optimal constructive
gales and whether constructive dimension can be equivalently defined
using constructive gales.  This paper addresses these questions.
\end{remark}

\section{The Strength of Gales}

\begin{theorem}\label{th:supergale_gale}Let $0 \leq r < t$ be
computable real numbers.  Then for any constructive $r$-supergale $d$,
there exists a constructive $t$-gale $d'$ such that $S^\infty[d]
\subseteq S^\infty[d']$.
\end{theorem}

\begin{proof}
Let $d$ be a constructive $r$-supergale and assume without loss of
generality that $d(\lambda) < 1$.  Define the language $A = \{ w \in
\{0,1\}^* | d(w) > 1 \}$.  Observe that $A$ is computably enumerable.
For all $n \in \N$, $\sum_{w\in\{0,1\}^n}d(w) \leq 2^{rn}$, so
$|A_{=n}| \leq 2^{rn}$.

For each $n \in \N$, define a
function $d_n' : \{0,1\}^* \to [0,\infty)$ by
\begin{equation*}
d_n'(w) = 
\begin{cases}
2^{-t(n-|w|)} \cdot \big|\{v \in A_{=n}| w \prefix v \}\big| & \textrm{if
$|w| \leq n$} \\ 
 2^{(t-1)(|w|-n)}d_n'(w[0..n-1]) & \textrm{if $|w| > n$.}
\end{cases}
\end{equation*}
Then for all $n$, $d_n'$ is a $t$-gale and $d_n'(w) = 1$ for all $w
\in A_{=n}$.

Let $s \in (r,t)$ be computable and define a function $d'$ on
$\{0,1\}^*$ by $d' = \sum_{n=0}^\infty 2^{(s-r)n}d_n'.$ Then
$$d'(\lambda) = \sum_{n=0}^\infty 2^{(s-r)n} 2^{-tn} |A_{=n}| \leq
\sum_{n=0}^\infty 2^{(s-t)n} < \infty,$$ and it follows that by induction that $d'(w) <
\infty$ for all strings $w$.  Therefore, by linearity, $d'$ is a
$t$-gale.  Also, because the language $A$ is computably enumerable,
$d'$ is constructive.

Let $S \in S^\infty[d]$.  Then for infinitely many $n\in\N$, $S[0..n-1] \in
A$.  For each of these $n$, $$d'(S[0..n-1]) \geq 2^{(s-r)n}
d_n'(S[0..n-1]) = 2^{(s-r)n},$$ so $S \in S^\infty[d']$.  Therefore
$S^\infty[d] \subseteq S^\infty[d']$.
\end{proof}

Constructive dimension may now be equivalently defined using gales
instead of supergales.

\begin{theorem}For all $X \subseteq \C$, $\cdim(X) = \inf \G_\constr(X)$.
\end{theorem}
\begin{proof}
Because any gale is also a supergale, $\G_\constr(X) \subseteq
\Ghat_\constr(X)$, so $\cdim(X) = \inf \Ghat_\constr(X) \leq \inf
\G_\constr(X)$ is immediate.

Let $t > r > \cdim(X)$ be computable real numbers and let $d$ be a
constructive $r$-supergale such that $X \subseteq S^\infty[d]$.  By
Theorem \ref{th:supergale_gale}, there is a constructive $t$-gale $d'$
such that $X \subseteq S^\infty[d] \subseteq S^\infty[d']$, so $t \in
\G_\constr(X)$.  As this holds for any computable $t > \cdim(X)$, we
have $\inf \G_\constr(X) \leq \cdim(X)$.
\end{proof}

We can also state the existence of a form of optimal constructive
gales.

\begin{corollary}\label{co:optimal_gales}
For all computable real numbers $t > r \geq 0$ there exists a constructive
$t$-gale that is successively optimal for the class of constructive
$r$-supergales.
\end{corollary}
\begin{proof}
Let $\bd^{(r)}$ be the constructive $r$-supergale from Theorem
\ref{th:lutz_optimal} that is multiplicatively optimal for the
constructive $r$-supergales.  Theorem \ref{th:supergale_gale} provides
a constructive $t$-gale $d'$ that succeeds everywhere that $\bd^{(r)}$
does.  Therefore $S^\infty[d] \subseteq S^\infty[\bd^{(r)}] \subseteq
S^\infty[d']$ for any constructive $r$-supergale $d$, so the corollary
is proved.
\end{proof}

The optimal gales provided by Corollary \ref{co:optimal_gales} 
may not be technically strong as possible in two respects.
\begin{enumerate}
\item Lutz's optimal constructive $r$-supergale is multiplicatively
optimal, whereas our optimal constructive $t$-gale is only
successively optimal.  Does there exist a constructive $t$-gale that
is multiplicatively optimal for the class of constructive
$r$-supergales?
\item Our proof seems to require the hypothesis $t > r$.  Does there
exist a constructive $r$-gale that is successively optimal for the class of
constructive $r$-supergales?
\end{enumerate}
However, the optimality in Corollary \ref{co:optimal_gales} remains strong
enough to prove Theorem \ref{th:lutz_pointwise}.

\begin{ack}I thank Anumodh Abey for comments on an earlier draft.\end{ack}

\bibliographystyle{plain}
\bibliography{\INC master}

\begin{thebibliography}{1}

\bibitem{Hitchcock:CPED}
J.~M. Hitchcock.
\newblock Correspondence principles for effective dimensions.
\newblock In {\em Proceedings of the 29th International Colloquium on Automata,
  Languages, and Programming}, pages 561--571, 2002.

\bibitem{Lutz:AEHNC}
J.~H. Lutz.
\newblock Almost everywhere high nonuniform complexity.
\newblock {\em Journal of Computer and System Sciences}, 44:220--258, 1992.

\bibitem{Lutz:DCC}
J.~H. Lutz.
\newblock Dimension in complexity classes.
\newblock In {\em Proceedings of the Fifteenth Annual IEEE Conference on
  Computational Complexity}, pages 158--169. IEEE Computer Society Press, 2000.
\newblock Updated version appears as Technical report cs.CC/0203016, ACM
  Computing Research Repository, 2002.

\bibitem{Lutz:GCDIS}
J.~H. Lutz.
\newblock Gales and the constructive dimension of individual sequences.
\newblock In {\em Proceedings of the Twenty-Seventh International Colloquium on
  Automata, Languages, and Programming}, pages 902--913. Springer-Verlag, 2000.

\bibitem{Lutz:DISS}
J.~H. Lutz.
\newblock The dimensions of individual strings and sequences.
\newblock Technical Report cs.CC/0203017, ACM Computing Research Repository,
  2002.

\bibitem{MartinLof66}
P.~Mar{tin-L{\"o}f}.
\newblock The definition of random sequences.
\newblock {\em Information and Control}, 9:602--619, 1966.

\bibitem{Schn71a}
C.~P. Schnorr.
\newblock A unified approach to the definition of random sequences.
\newblock {\em Mathematical Systems Theory}, 5:246--258, 1971.

\bibitem{Schn71b}
C.~P. Schnorr.
\newblock {Zuf\"alligkeit} und {Wahrscheinlichkeit}.
\newblock {\em Lecture Notes in Mathematics}, 218, 1971.

\bibitem{ZvoLev70}
A.~K. Zvonkin and L.~A. Levin.
\newblock The complexity of finite objects and the development of the concepts
  of information and randomness by means of the theory of algorithms.
\newblock {\em Russian Mathematical Surveys}, 25:83--124, 1970.

\end{thebibliography}

\end{document}